\documentclass[12pt]{article}
\usepackage[super,compress]{cite}
\usepackage{graphicx}
\usepackage{epsfig}
\usepackage{amssymb}
\begin{document}

\begin{center}
{\bf Central and peripheral hadron-nucleus collisions\\
in the Additive Quark Model
}

\vspace{.2cm}

G.H. Arakelyan$^1$,
Yu.M. Shabelski$^2$ and A.G. Shuvaev$^2$ \\

\vspace{.5cm}

$^1$A.Alikhanyan National Scientific Laboratory
(Yerevan Physics Institute)\\
Yerevan, 0036, Armenia\\
E-mail: argev@mail.yerphi.am\\

\vspace{.2cm}

$^1$Petersburg Nuclear Physics Institute, Kurchatov National
Research Center\\
Gatchina, St. Petersburg 188300, Russia\\
\vskip 0.9 truecm
E-mail: shabelsk@thd.pnpi.spb.ru\\
E-mail: shuvaev@thd.pnpi.spb.ru

\vspace{1.2cm}

\end{center}

\begin{abstract}
\noindent

Peripheral nucleon-nucleus collisions occur
at the high energies mainly through the interaction
with one constituent quark from the incident nucleon.
The central collisions should involve all three
constituent quarks and each of them can interact
several times. We calculate the average
number of quark-nucleus interactions for both
the cases in good agreement with the experimental
data on $\phi$-meson, $K^{*0}$ and all charged
secondaries productions in $p+Pb$ collisions
at LHC energy $\sqrt s = 5$~TeV.
\end{abstract}

\section{Introduction}

The Additive Quark Model (AQM) treats the nucleon
as a system of three quasi free
constituent quarks~\cite{LF, Lipkin:1965fu}.
They play the roles of incident particles
in terms of which the hadron scattering is described.
First phenomenological predictions of AQM
\cite{Anisovich:1977yk, Anisovich:1981bv, Anisovich:1985xd}
have demonstrated a fairly good agreement with experimental
data.

Basically, the scattering amplitude is presented in AQM
as a sum over the terms with a given number of constituent
quarks involved into the process.
For the proton-nucleus scattering there are three
classes of interactions depending on whether only one,
two or three incident constituent quarks
participate~\cite{Anisovich:1977yk, Anisovich:1985xd, Anisovich:1977av}.
The sum of their probabilities is normalized to unity.

The probabilities of all three types of events
are of the same order for heavy nuclear target
at the fixed target energies~\cite{Anisovich:1977av}.
The probability of the three quark interaction
grows at the LHC energies
because of the growth of the interaction cross section
thereby reducing two other probabilities
due to the total normalization.

All three events are essentially dependent
on the hadron-nucleus impact parameter.
Three quark interaction dominates in the central collisions.
The configuration with two interacting and one non interacting
quarks has a smaller probability which in turn is larger
than the probability to have a single interacting quark only.
The situation for the peripheral hadron-nucleus
collisions is opposite. Here the one quark interaction
is most probable while the probability of the three quark
interaction is minimal. Thus the peripheral hadron-nucleus
scattering looks more close to the hadron-nucleon one.
As a result the multiplicity of the produced
secondaries in the central collisions
should be about three times larger larger than for
the peripheral collisions.

In the present paper we calculate the total number
of the incident constituent quarks interactions
with the target nucleons for the all three event
classes. The ratios of the numbers of
the quark interactions in
central and peripheral collisions is compared
with the LHC experimental
data at $\sqrt s = 5$~TeV \cite{Adam:2015vsf,Adam:2016bpr}.

\section{Calculation of the probabilities and the number
of quark interactions for various event classes.}

The probabilities for the one, two and all three quarks
from the fast nucleon to interact with the target nucleus
were firstly calculated in ref.~\cite{Anisovich:1977av}.
They have the form
\begin{eqnarray}
v_1\,&=&\,\frac 3{\sigma_{pA}^{inel}}
\int d^{\,2}b\,e^{-2\sigma_q T(b)}
\bigl(1-e^{-\sigma_q T(b)}\bigr) \nonumber \\
\label{v1v2v3}
v_2\,&=&\,\frac 3{\sigma_{pA}^{inel}}
\int d^{\,2}b\,e^{-\sigma_q T(b)}
\bigl(1-e^{-\sigma_q T(b)}\bigr)^2  \\
v_3\,&=&\,\frac 1{\sigma_{pA}^{inel}}
\int d^{\,2}b\,
\bigl(1-e^{-\sigma_q T(b)}\bigr)^3. \nonumber
\end{eqnarray}
Here the target nuclear profile function,
$$
T(b)\,=\,A\,\int_{-\infty}^\infty \rho(b,z) dz,
$$
is given by Fermi nuclear matter distribution
$$
\rho(r)\,=\,\frac{\rho_0}{1 + e^{\frac{r-c_1}{c_2}}}
$$
with the parameters
$$
c_1\,=\,1.15 A^{1/3}~{\rm fm},~~~~
c_2\,=\,0.51~{\rm fm}
$$
and $\rho_0$ value determined by the normalization
$\int d^3r \rho(r) = 1$.
The cross section of the proton-nucleus
inelastic scattering,
\begin{equation}
\label{pA}
\sigma_{pA}^{inel}\,=\,\int d^{\,2}b\,
\bigl(1-e^{-\sigma_{pN}^{inel}\, T(b)}\bigr),
\end{equation}
is expressed through
the proton-nucleon inelastic cross-section
$\sigma_{pN}$. The constituent quark inelastic
cross-section with the target nucleon is
assumed to be 1/3 of the proton-nucleon one,
$\sigma_q =1/3\,\sigma_{pN} ^{inel}$.

Given the factor $e^{-\sigma_q T(b)}$ as a probability
for a quark not to interact with the target at the distance $b$
the expressions (\ref{v1v2v3}) are rather evident.
The value $v_1$ stands for the processes
where one of the three quarks
interacts with the nucleus while the other two quarks do not.
The $v_2$ value refers to the opposite situation,
$v_3$ is the probability
for all three quarks to interact,
$v_1+v_2+v_3=1$.

The cross section (\ref{pA}) can be recast in the form
\begin{eqnarray}
\sigma_{pA}^{inel}\,&=&\,\int d^{\,2}b\,
e^{-\sigma_{pN}^{inel}\, T(b)}\,
\bigl[e^{\sigma_{pN}^{inel}\, T(b)}-1\bigr]\nonumber \\
&=&\,
\sum_{\nu=1}^\infty \int d^{\,2}b\,
e^{-\sigma_{pN}^{inel}\, T(b)}
\frac{1}{\nu!}\,\bigl[\,\sigma_{pN}^{inel}\, T(b)\,\bigr]^\nu
\,=\,\sum_{\nu=1}^\infty \sigma_{pA}^{(\nu)}.\nonumber
\end{eqnarray}
Each term comes to the sum from the interactions with $\nu$
target nucleons. The average number of the collisions
in the proton-nucleus scattering is therefore equal to
\begin{eqnarray}
\label{nupA}
\langle\,\nu\,\rangle_{pA}\,&=&\,\frac 1{\sigma_{pA}^{inel}}
\sum_{\nu=1}^\infty \nu\,\sigma_{pA}^{(\nu)}\\
&=&\,\frac 1{\sigma_{pA}^{inel}}
\int d^{\,2}b\,e^{-\sigma_{pN}^{inel}\, T(b)}
\sum_{\nu=1}^\infty \frac{1}{(\nu-1)!}\,
\bigl[\,\sigma_{pN}^{inel}\, T(b)\,\bigr]^\nu\,=\,
A\,\frac {\sigma_{pN}^{inel}}{\sigma_{pA}^{inel}}.
\nonumber
\end{eqnarray}

To find the inclusive density of the secondaries
in the central (midrapidity) region
it is necessary to calculate the average
number of collisions with the target
nucleus.
This number gets separate contributions
from the three classes of events specified by
equations (\ref{v1v2v3}).
Taking the probability for the first class,
$$
v_1\,=\,\frac 3{\sigma_{pA}^{inel}}
\int d^{\,2}b\,e^{-3\sigma_q T(b)}
\bigl(e^{\sigma_q T(b)}-1\bigr),
$$
the relevant collisions number $\langle\nu_1\rangle$
is obtained similar to Eq.(\ref{nupA}),
\begin{eqnarray}
v_1\,\cdot\,\langle \nu_1\rangle\,&=&\,
\frac 3{\sigma_{pA}^{inel}}
\int d^{\,2}b\,e^{-3\sigma_q T(b)}
\sum_{\nu=1}^\infty
\nu\, \frac{[\sigma_q T(b)]^\nu}{\nu!}
\nonumber \\
\label{v1nu1}
&=&\,\frac 3{\sigma_{pA}^{inel}}
\int d^{\,2}b\,e^{-2\sigma_q T(b)}
\sigma_q T(b).
\end{eqnarray}
The second class probability,
$$
v_2\,=\,\frac 3{\sigma_{pA}^{inel}}
\int d^{\,2}b\,e^{-2\sigma_q T(b)}
\bigl(e^{\sigma_q T(b)}-1\bigr)^2,
$$
is worked out in the same manner.
The parenthesis is expanded and
the terms with $[\sigma_q T(b)]^\nu$
are regarded as arising from $\nu$ collisions.
Multiplying them by $\nu$ and summing up
the series one gets
\begin{equation}
\label{v2nu2}
v_2\,\cdot\,\langle \nu_2\rangle\,=\,
\frac 6{\sigma_{pA}^{inel}}
\int d^{\,2}b\,e^{-2\sigma_q T(b)}
\bigl(e^{\sigma_q T(b)}-1\bigr)\,
\sigma_q T(b).
\end{equation}
Repeating these steps for the third class,
$$
v_3\,=\,\frac 1{\sigma_{pA}^{inel}}
\int d^{\,2}b\,e^{-3\sigma_q T(b)}
\bigl(e^{\sigma_q T(b)}-1\bigr)^3,
$$
returns the value
\begin{equation}
\label{v3nu3}
v_3\,\cdot\,\langle \nu_3\rangle\,=\,
\frac 3{\sigma_{pA}^{inel}}
\int d^2{\,b}\,e^{-2\sigma_q T(b)}
\bigl(e^{\sigma_q T(b)}-1\bigr)^2\,
\sigma_q T(b).
\end{equation}
All three classes yield in aggregate
the average proton-nucleus collision number,
$$
\nu_{pA}\,=\,
v_1\,\cdot\,\langle \nu_1\rangle\,+\,
v_2\,\cdot\,\langle \nu_2\rangle\,+\,
v_3\,\cdot\,\langle \nu_3\rangle\,=
\frac 1{\sigma_{pA}^{inel}}
\int d^{\,2}b\,\sigma_{pN}T(b)\,=\,
A\frac{\sigma_{pN}}{\sigma_{pA}^{inel}}.
$$

The equation (\ref{nupA}) gives
for the proton-lead collision
at the LHC energy $\sqrt s=5$~TeV
(taking $\sigma_{pN}^{inel} = 69.86$~mb
and $\sigma_{pA}^{inel} =1965$~mb
\cite{Shabelski:2016aek})
\begin{equation}
\langle\,\nu\,\rangle_{pA}\,=\,7.36.
\end{equation}

When only one incident constituent quark
interacts with the lead nucleus one obtains
from the equations (\ref{v1v2v3}) and (\ref{v1nu1})
$v_1=0.19$, $v_1\cdot\langle\nu_1\rangle=0.26$,
while for the two or three interacting quarks
the equations (\ref{v1v2v3}) and (\ref{v2nu2}),(\ref{v3nu3})
give $v_2= 0.20$, $v_3= 0.61$,
$v_2\cdot\langle\nu_2\rangle = 0.82$,
$v_3\cdot\langle\nu_3\rangle = 6.297$,
that means
$$
\langle\nu_1\rangle\,=\, 1.28,~~~
\langle\nu_2\rangle\,=\,4.16,~~~
\langle\nu_3\rangle\, =\,10.31.
$$

\section{Numerical results and conclusion}

As has been mentioned above $\nu_1$ is related to
the peripheral proton-lead collisions, $\nu_3$ to the central
collisions and $\nu_2$ to the intermediate type.
The experimental data on the inclusive secondaries
density in the midrapidity region are presented in
refs.~\cite{Adam:2015vsf, Adam:2016bpr}
for $p+Pb$ scattering at $\sqrt s=5$~TeV.
In these papers the events are divided into several
classes with respect to the mean multiplicities.
It is reasonable to assume that the smallest
and the highest multiplicities refer to
the peripheral and the central collisions
whereas the intermediate interactions occur mainly
in the collisions of two constituent quarks.
However the intermediate region should
include also the tails from the single and three quarks
interactions that makes it more complicated to analyze.
The outcome should be somewhere between the two limiting cases
but we do not consider it.

The ratio of the inclusive densities of any secondaries
in the central and peripheral collisions is compared below
with the obtained $\nu_3/\nu_1$ ratio.

\medskip
Table.
Comparison of the calculated $\nu_3/\nu_1$ values
with the experimental data for the ratio of
the inclusive densities in the central (event classes
0--20\%) and peripheral (event classes 80--100\%)
collisions for the all charged secondaries,
$K^{*0}$ and $\phi$
(see refs.~\cite{Adam:2015vsf,Adam:2016bpr}
for details).

\begin{center}
\begin{tabular}{|c|c|c|c|}\hline
$\nu_3/\nu_1$&
\multicolumn{3}{|c|}
{$\frac{dN}{dy}$(central)$/\frac{dN}{dy}$(peripheral)}\\
\hline
& all charged& $K^{*0}$& $\phi$ \\
\cline{2-4}
8.05 &         $8.3 \pm 0.4$ & $7.4 \pm 1.2$ & $9.7 \pm 1.4$ \\
\hline
\end{tabular}
\end{center}
\medskip

The presented table shows unexpectedly good
agreement of our calculations with
the experimental data.
The important point is that the experimental
ratios for the different
secondaries coincide inside the error bars.
The theoretical uncertainties can be estimated
at the level 10-15\%, this level is confirmed
by the validity of another AQM predictions \cite{Anisovich:1981bv}.

It will be interesting to investigate the same ratios
for another secondary particles
as well as another nuclear targets.

\end{document}